\input harvmac.tex
%

%
%

\def\IB{\relax\hbox{$\inbar\kern-.3em{\rm B}$}}
\def\IC{\relax\hbox{$\inbar\kern-.3em{\rm C}$}}
\def\ID{\relax\hbox{$\inbar\kern-.3em{\rm D}$}}
\def\IE{\relax\hbox{$\inbar\kern-.3em{\rm E}$}}
\def\IF{\relax\hbox{$\inbar\kern-.3em{\rm F}$}}
\def\IG{\relax\hbox{$\inbar\kern-.3em{\rm G}$}}
\def\IGa{\relax\hbox{${\rm I}\kern-.18em\Gamma$}}
\def\IH{\relax{\rm I\kern-.18em H}}
\def\IK{\relax{\rm I\kern-.18em K}}
\def\IL{\relax{\rm I\kern-.18em L}}
\def\IP{\relax{\rm I\kern-.18em P}}
\def\IR{\relax{\rm I\kern-.18em R}}
\def\IZ{\relax\ifmmode\mathchoice
{\hbox{\cmss Z\kern-.4em Z}}{\hbox{\cmss Z\kern-.4em Z}}
{\lower.9pt\hbox{\cmsss Z\kern-.4em Z}}
{\lower1.2pt\hbox{\cmsss Z\kern-.4em Z}}\else{\cmss Z\kern-.4emZ}\fi}

\def\CC {{\cal C}}


\def\p{\partial}



\def\Tr{{\rm Tr}}

\def\p{\partial}

\def\np{\nabla_{\partial}}

\def\inbar{\,\vrule height1.5ex width.4pt depth0pt}
\font\cmss=cmss10 \font\cmsss=cmss10 at 7pt

\def\O{\Omega}
\def\o{\omega}

\def\D{\Delta}

\def\d{\delta}
\def\e{\epsilon}
\def\m\mu
\def\n{\nu}

\def\p{\partial}
\def\R{\relax{\rm I\kern-.18em R}}
\font\cmss=cmss10 \font\cmsss=cmss10 at 7pt
\def\Z{\relax\ifmmode\mathchoice
{\hbox{\cmss Z\kern-.4em Z}}{\hbox{\cmss Z\kern-.4em Z}}
{\lower.9pt\hbox{\cmsss Z\kern-.4em Z}}
{\lower1.2pt\hbox{\cmsss Z\kern-.4em Z}}\else{\cmss Z\kern-.4em Z}\fi}
\def\pl{{\it  Phys. Lett.}}

\def\np{{\it Nucl. Phys. }}

\def\hf{{1\over 2}}

\def\ra{\rightarrow}

\def\NI{{1\over N}}
\def\dN{d^{N^2}\!\!}
 %
  %


\lref\DSL{  E.~Br{\'e}zin, V.~A.~Kazakov, Exactly solvable field theories of
        closed strings, Phys. Lett. B236 (1990) 144;    

M.~R.~Douglas, S. Shenker, Strings in less than one dimension,
Nucl.Phys. B335 (1990) 635;

D.~Gross, A.~Migdal, Non-perturbative two-dimensional quantum gravity, 
 Phys.Rev.Lett. 64 (1990) 127.    }

\lref\KSWI{ V.A. Kazakov, M. Staudacher and T. Wynter, Exact Solution 
of Discrete Two-Dimensional $R^2$ Gravity, hep-th/9601069,
Nucl.Phys.B471 (1996) 309-333.  }

\lref\KSWII{V.A. Kazakov, M. Staudacher and  T. Wynter, Character expansion
                    methods for matrix models of dually weighted
                    graphs, hep-th/9502132, Comm. Math. Phys. 177
                    (1996) 451-468.}

\lref\KSWIII{ V.A. Kazakov, M. Staudacher, T. Wynter, 
Almost flat planar graphs, hep-th/9506174, Comm. Math.Phys. 179 (1996)
235-256.  }

\lref\HKK{ J.~Hoppe, V.~Kazakov and I.~Kostov,
Dimensionally Reduced $SYM_4$ as Solvable Matrix Quantum
Mechanics, hep-th/9907058, to be published in \np B. }

\lref\KKN{V. Kazakov, I. Kostov, N. Nekrasov, D-particles, Matrix
Integrals and KP hierachy, hep-th/9810035,
\np B557 (1999) 413-442.}

\lref\DK{M. R. Douglas and V. A. Kazakov, Large $N$
Phase Transition in Continuum $QCD_2$,
\pl B319 (1993) 219 hep-th/9305047. }

\lref\KAZBU{D. Boulatov and V. Kazakov, One dimensional string theory
with vortices as an upside down matrix oscillator, {\it J.Mod.Phys}
A8 (1993) 809.}

\lref\KazMig{ V. Kazakov and A. Migdal, Recent Progress in the Theory of
Non-critical Strings, \np B 311 (1988) 171.}

\lref\CC{  Charich-Chandra, Amer. J. Math., 79 (1957) 87.  }

\lref\IZ{C.Itzykson and J.B.Zuber, Planar approximation II,  
J.Math.Phys.  21 (1980) 411.  }

\lref\KLGR{ D.Gross and I.Klebanov, Vortices and the non-singlet 
sector of the c=1 matrix model, \np B354 (1990) 459.}

\lref\MARON{G.~Marchesini and E.~Onofri, Planar limit for  $SU(N)$ 
symmetric quantum dynamical systems, J.Math.Phys.  21 (1980) 1103.}

\lref\BIPZ{ E.~Brezin,  C.~Itzykson,  G.~Parisi and J.-B.~Zuber,   
Planar approximation,  Comm. Math. Phys. 59 (1978) 35. }

\lref\CHMM{ S.~Chadha, G.~Mahoux and M.~L.~Mehta, 
A method of integration over matrix variables 2, J. Phys. A:
Math. Gen.  14 (1981) 579.  }

\lref\MEHTA{ M.~L.~Mehta, A method of integration over matrix variables, 
Comm. Math. Phys. 79 (1981) 327. }

\lref\ISING{  V.~A.~Kazakov, Ising model on dynamical planar random lattice:
          exact solution, Phys. Lett. 119A (1986) 140;  

 D.~V.~Boulatov and V.~A.~Kazakov, The Ising model on a random planar
          lattice : the structure of phase transition and the exact
          critical exponents, Phys.Lett. B186 (1987) 379.}

\lref\Kaz{ V.~A.~Kazakov, D-dimensional induced gauge theory as a solvable
                    matrix model, Proc. Intern. Symp. "Lattice '92"
                    on Lattice gauge theory, Amsterdam, Sept. 1992,
                    Nucl. Phys.  B(Proc. Suppl.) 30 (1993) 149.}

\lref\DOUG{ M.~R.~Douglas, Strings in less than one dimention and 
generalized KdV hierarchies, Phys.Lett. 238B (1990) 176.}

\lref\KOS{ I.~K.~Kostov, Gauge Invariant Matrix Model for the ADE Closed 
Strings, hep-th/9208053, Phys.Lett. B297 (1992) 74-81. }

\lref\KAZP{  V.~A.~Kazakov, Exactly solvable Potts models, bond and tree-like
          percolation on dynamical (random) planar lattice,
          Nucl. Phys. B 4 (Proc. Supp.) (1988) 93.  }

\lref\KAKO{  V.~A.~Kazakov and  I.~K.~Kostov, published in the review of 
I.~K.~Kostov, Random surfaces, solvable matrix models and discrete
quantum gravity in two dimensions, Lecture given at GIFT Int. Seminar
on Nonperturbative Aspects of the Standard Model, Jaca, Spain, Jun
6-11, 1988. Published in GIFT Seminar 0295 (1988) 322.}

\lref\DAUL{  J-M.~DAUL, Q-states Potts model on a random planar lattice,
     hep-th/9502014. }

\lref\EYNARD{
B.~Eynard, G.~Bonnet, The Potts-q random matrix model : loop
     equations, critical exponents, and rational case, Phys.Lett. B463
     (1999) 273-279.}

\lref\KAZDO{  V.~A.~Kazakov,  Bosonic strings and string field theories in
                    one-dimensional target space, in "Random surfaces
                    and quantum gravity", Carg se 1990, O. Alvarez,
                    E. Marinari, P. Windey eds., (1991). }

\lref\MIG{A.~A.~Migdal, Recursion equations in lattice gauge theories, 
Sov.Phys.JETP 42 (1975) 413, Zh.Eksp.Teor.Fiz.69 (1975) 810-822. }

\lref\DIFZ{ Ph.~DiFrancesco and C.~Itzykson, Fat graphs, Ann. Inst. 
Henri Poincar\`e, 59(2) (1993) 117.  }

\lref\KSTW{I. K. Kostov, M. Staudacher, T. Wynter,  
Complex Matrix Models and Statistics of Branched Coverings of 2D Surfaces,
hep-th/9703189, Comm.Math.Phys. 191 (1998) 283-298.}

\lref\KZJ{V. A. Kazakov, P. Zinn-Justin, 
Two-Matrix model with ABAB interaction, hep-th/9808043,
Nucl.Phys. B546 (1999) 647-668. }

\lref\HKK{J. Hoppe, V. A. Kazakov, I. K. Kostov, Dimensionally Reduced 
$SYM_4$ as Solvable Matrix Quantum Mechanics, hep-th/9907058,
Nucl.Phys.B (to be published). }

\lref\KKN{ V. A. Kazakov, I. K. Kostov, N. Nekrasov, D-particles, 
Matrix Integrals and KP hierarchy, hep-th/9810035, Nucl.Phys. B557
(1999) 413-442.}

\lref\POLY{ A. P. Polychronakos, Generalized statistics in one 
dimension, hep-th/9902157, Les Houches 1998 Lectures; 54 pages. }

\lref\MMM{ S. Kharchev, A. Marshakov, A. Mironov, A. Morozov,
Generalized Kontsevich Model Versus Toda Hierarchy and Discrete Matrix Models,
hep-th/9203043, Nucl. Phys. B397 (1993) 339. }

\lref\MOR{ S. Kharchev, A. Marshakov,
     A. Mironov, A. Morozov, S. Pakuliak, Conformal Matrix Models as
an Alternative to Conventional Multi-Matrix Models, hep-th/9208044,
Nucl. Phys. B404 (1993) 717.  }

\lref\KS{I. K. Kostov and M. Staudacher,  
Two-Dimensional Chiral Matrix Models and String Theories,
hep-th/9611011, Phys.Lett. B394 (1997) 75-81. }

\lref\KOSR{I. K. Kostov, Bilinear Functional Equations in 2D Quantum Gravity,
  hep-th/9602117, Talk delivered at the Workshop on "New Trends in
     Quantum Field Theory", 28 August - 1 September 1995, Razlog,
     Bulgaria.  }


\Title{}
{\vbox{
\centerline{  Solvable Matrix Models}
 \vskip2pt
}}
%
%
%
\centerline{Vladimir Kazakov \footnote{$ ^\bullet $}{{\tt
kazakov@physique.ens.fr}}}
\centerline{{\it $^1$  Laboratoire de Physique Th\'eorique de l'Ecole
Normale Sup\'erieure \footnote{$ ^\ast $}{
Unit\'e Mixte du
Centre National de la Recherche Scientifique
et de l'Ecole Normale Sup\'erieure.}}}

\centerline{{\ \ \ \it  75231 Paris CEDEX, France}}

 \vskip 1cm
\baselineskip8pt{
 
\vskip .2in
 
\baselineskip8pt{

 We review some old and new methods of reduction of the number of degrees
of freedom from $\sim N^2$ to $\sim N$ in the multi-matrix integrals.

{\it A talk delivered at the MSRI Workshop  ``Matrix Models and
Painlev\'e Equations'' , Berkeley (USA) 1999 }

\bigskip
 
\rightline{ LPTENS-00/09}

\Date{February, 2000}


\baselineskip=20pt plus 2pt minus 2pt

\newsec{Introduction }

Multi-matrix integrals of various types appear in many
mathematical and physical applications, such as combinatorics of
graphs, topology, integrable systems, string theory, theory of
mesoscopic systems or statistical mechanics on random surfaces.

A general Q-matrix integral of the form
 \eqn\mmm{ Z=\int\prod_{q=1}^Q \dN M_q\exp S(M_1,\cdots,M_Q) }
usually goes over the $N\times N$ hermitian, real symmetric or
symplectic matrices $M_q$ with the action $S$ and the measure
symmetric under the simultaneous group rotation:
$M_q\ra\O^+M_q\O$. Some other multi-matrix integrals, such as these
with complex matrices or with general real matrices, can be reduces to
those three basic cases.
 
We will consider here only the case of hermitean matrices for which
$\O$ belongs to the $U(N)$-group.

In many applications "to solve" the corresponding matrix model usually
means to reduce the number of variables by explicit integrations over
most of the variables in such a way that instead of $QN^2$ original
integrations (matrix elements) one would be left in the large $N$
limit only with $\sim N$ integration variables. In this case the
integration over the rest of the variables can be performed, at least
in the widely used large N limit, by means of the saddle point
approximation. A more sophisticated double scaling limit \DSL\ is also
possible (if possible at all) only after such a reduction. The key of
success is in the fact that after reduction the effective action at
the saddle point is still of the order $\sim N^2$ whereas the
corrections given by the logarithm of determinant of the second
variation of the action cannot be bigger than $\sim N$ (the "entropy"
of the remaining variables). The problem is thus reduced to the
solution of the "classical" saddle point equations, instead of the
"quantum" problem of functional (in the large N limit) integration
over the original matrix variables.

Such an explicit reduction of the number of "degrees of freedom" is
in general possible only for a few rather restricted, though
physically and mathematically interesting, classes of multi-matrix
integrals. The purpose of our present notes is to review the basic old
and new methods of such a reduction. Before going to the particular
cases let us stress the importance of the search for new methods of
such a reduction: any nontrivial finding on this way leads immediately
to numerous fruitful applications.

\newsec{ Some old examples}

 The best known example of such a reduction of the number of degrees of freedom is the one matrix integral:
 \eqn\onemm{ Z=\int \dN M\exp N\Tr S(M) }
where $S(M)$ is an arbitrary function of one variable. Let us use
the decomposition:
\eqn\dec{ M=\O^+ x \O }
where $x=diag(x_1,\cdots,x_N)$ is a diagonal matrix of the
eigenvalues and $\O$ is the $U(N)$ group variable. The
corresponding (Dyson) measure can be written as:
\eqn\meas{\dN M=d[\O]_{U(N)}\D^2(x) \prod_{k=1}^N dx_k  }
where $\D(x)=\prod_{i>j}(x_i-x_j)$ is the Van-der-Monde
determinant. The integrand as an invariant function does not depend at
all on $\O$ (the integration over it produces just a group volume
factor which we will always omit). The remaining integral over the
eigenvalues reads:
 \eqn\oneme{ Z=\int\prod_{k=1}^N dx_k\exp[N S(x_k)] \D^2(x)}
In the large N limit the corresponding saddle point equation
takes the form
\eqn\spe{ \NI {\p S\over \p x_k}=S'(x_k)+{1\over N}
\sum_{j\ne k} {1\over x_k-x_j}=0 }
These arguments were successfully used for an interesting
combinatorial problem: enumeration of graphs of fixed two dimensional
topologies \BIPZ , \IZ . There exist powerful methods to analyze this
equation but it is not our present goal to review them here.

The next fruitful example is the so called two matrix model:
 \eqn\twomm{ Z=\int \dN A \ \dN B\exp N\Tr
 \big(-A^2-B^2+cAB+U(A)+ V(B)\big)}
where $U$ and $V$ are some arbitrary functions of one variable.
After the decomposition $A=\O_1^+x\O_2$, $B=\O_2^+y\O_2$ we are
left, due to the term $\Tr(AB)$ in the action, with one
nontrivial unitary integral over the variable $\O=\O_1\O_2^+$.
Fortunately, this integral was explicitly calculated by
Charish-Chandra, Itzykson and Zuber \CC , \IZ :
 \eqn\CCIZ{ \int d[\O]_{U(N)} \exp\Tr(\O^+x\O y)=\prod_{k=1}^{N-1}k!
{\det_{ij}e^{x_iy_j}\over \D(x)\D(y)} }
Substituting \CCIZ\ and the Dyson measure \meas\ into \twomm\
we are left again with only $2N$ variables $x_k$ and $y_k$ and we
can write again the saddle point equations in the large N limit.
They are more complicated than in the one matrix integral but can
be nevertheless solved quite explicitly. The first solution of
that kind was found in \MEHTA\ in an indirect way, using the
method of orthogonal polynomials, but the direct solution is also
possible (see \Kaz ).

This model was used in \ISING\ to solve exactly the first example of new
statistical mechanical models of interacting spins on random planar
graphs: in this case it was a model of Ising spins on random planar
graphs.

An obvious generalization of the two matrix model is the  matrix chain model:
 \eqn\MMM{ Z=\int\prod_{q=1}^Q \dN M_q \ \ \exp \Tr\big(\sum_{p=1}^Q
V_q(M_q) +\sum_{p=1}^{Q-1} M_{p-1}M_p \big)}
One easily notices that the same unitary decomposition
$M_q=\O_q^+x_q\O_q$ leads to $Q-1$ independent integrals over the
variables $U_q=\O_{q-1}^+\O_q$ of the type \CCIZ . We are left again
with only $QN$ eigenvalues instead of $QN^2$ matrix elements and are
ready to apply the saddle point approximation to this integral. This
model was first analyzed by the method of orthogonal polynomials by
\CHMM . It was shown in \DOUG\ that by special choices of the
potential $V$ the model can be described by the KP integrable flow
with respect to the coupling constant of the potential.

Note that if we imposed the periodicity condition $M_1=M_q$ on this
matrix chain and add the term $M_1 M_q$ to the action the problem
would become much more complicated (and actually not solved so far),
since this would give an extra condition $\prod_q U_q=I$ making the
variables $U_q$ not independent.

Another solvable matrix chain  describing the statistical RSOS
RSOS models on random planar graphs was proposed and solved
in \KOS . Similar models were considered in  \MOR .

Some multi-matrix models can be reduced to the solvable ones by means
of simple matrix integral transformations. The first example of such
transformation was described in the paper \KAZP\ for the matrix
integral describing the Q-state Potts model on random dynamical planar
graphs. Its partition function is
 \eqn\PMM{ Z=\int\prod_{q=1}^Q \dN M_q\exp \Tr\big
( \sum_{q=1}^Q V_q(M_q) +\sum_{p,q=1}^{Q} M_pM_q \big)}
 One can represent the last factor under the integral as
$$\int \dN X \exp \Tr\big( -\hf X^2 +X\sum_{q=1}^Q M_q\big).$$
Let us consider the case $V_1=\cdots=V_Q=V$. Then the whole integral
can be expressed as
 \eqn\PMMX{ Z=\int \dN X \exp ( -\hf \Tr X^2)\Big[\int \dN M
 \exp \Tr\big( X M + V(M)\big)\Big]^Q  }
The integrals in \PMMX\ can be reduced to the eigenvalues: in the
integral under the power the only nontrivial ``angular'' integration
over the relative $U(N)$-''angle'' can be done by means of the
formula \CCIZ\ and the external one will also depend only on the
eigenvalues of $X$. The solution of the corresponding saddle point
equations was found in \KAKO\ and analyzed in \DAUL\ and \EYNARD .

Combining these methods in the obvious ways one can generalize the
large $N$ solvability on a certain larger class of multi-matrix
models.

\newsec{Matrix Quantum Mechanics}

In the limit when $Q\ra \infty$ and with the special scaling of
coupling constants the matrix chain \MMM\ becomes  matrix quantum
mechanics.  It is defined by the Hamiltonian
\eqn\MQM{ \hat H_M= -\D_M+\Tr V(M)    }
where $\D_M$ is the usual $U(N)$ invariant Laplacian on the
homogeneous space of  hermitian matrices and the potential
$V(M)$ can actually explicitly depend on time $t$.

The Schroedinger equation can be written in the form of a
minimization principle:
\eqn\SCHR{ min_\Psi \int \dN M \Tr\big( \hf|\p_M
\Psi(M)|^2+V(M)|\Psi(M)|^2) }
To reduce this problem to the eigenvalues we use the $U(N)$
symmetry of our model and look for a wave function $\Psi(M)$
transforming according to a certain irreducible representation
$R$ of $U(N)$: 
$$\Psi_R^I(\O^+M\O)=\sum_J\O^{IJ}_R\Psi^{J}_R(M)$$ 
where $\O_R$
is a group element  $\O$ in  representation $R$ and $I,J$ are the 
indices of the representation. Such a
function may be decomposed as
\eqn\TRANS{\Psi_R^I(M)=\sum_J\O_R^{IJ}\psi_R^J(x). }
Here $\psi_R^I(x_1,\cdots,x_N)$ is a vector in the
representation $R$.

Near the unity element on the group space $\O\simeq I +\omega$ we
have $\O_R\simeq P_R + \sum_{ij}\o_{ij}T^R_{ij}$ where $P_R$ is a
projector (unity element) in the $R$ space, $\o$ is a small deviation
from it and $T^R_{ij}$ are the $u(N)$ algebra generators. This gives:
$${\p\over \p M_{ij}}= \d_{kj} {\p\over \p x_k} +\sum_{m=1}^N{1\over
x_k-x_m}{\p\over \p\o_{mj}}$$ and we finally obtain from \SCHR\ the
following variational principle:
\eqn\SCHRXM{  {\rm min}_{\psi_R} \int \prod_k dx_k\D^2(x)
\Tr_R\big( \hf \sum_j|{\p \over \p x_j}\psi_R(x)|^2+
\hf\sum_{i\ne j} |T^R_{ij}\psi_R|^2+\sum_mV(x_m)|\psi_R|^2 \big) }
where all the quantities and operators with the subscript $R$ are
subjected to the corresponding matrix operations in the matrix space
of representation.

The Schroedinger equation now reads:
 \eqn\SCHRX{ -\sum_k\D^{-2}(x){\p\over \p x_k}\D^2(x){\p\over \p
x_k}\psi_R(x)-\sum_{i\ne j} T^R_{ij}T^R_{ji}\psi_R(x)=
\big(E-\sum_k V(x_k)\big)\psi_R(x) }
It is useful to introduce a new function $\phi_R(x)={1\over
\D(x)} \psi_R(x)$ obeying the equation
\eqn\SCHRXP{ -\sum_k\left({\p\over\p x_k}\right)^2\phi_R(x)-\sum_{i\ne j}
{T^R_{ij}T^R_{ji}\over (x_i-x_j)^2}\phi_R(x)=
\big(E-\sum_i V(x_i)\big)\phi_R(x) }
Note that any translation $\o_{ij} \ra \o_{ij}+ \d_{ij}\e $ does not
change the wave function $\Psi_R$. That means that we are looking only
for the states on which the condition is imposed
\eqn\COND{  T^R_{kk}\psi_R=0, \ \ k=1,\cdots,N }

On the first sight, we fulfilled our main task for the matrix
quantum mechanics: we reduced it to an eigenvalue problem and are
now dealing with only $N$ variables. But the Schroedinger
equation \SCHRX\ contains the Hamiltonian which is a matrix in
the representation space acting on the wave function which is a
vector in this space. For  small representations whose Young
tableaux contain $<<N^2$ boxes the problem is still solvable in
the large N limit (as we will demonstrate below). For a very
interesting case of big representations ($\sim N^2$ boxes in the
Young tableaux) the problem remains a serious challenge.

In the simplest case of singlet representation (solved long ago in
\BIPZ ) the wave function is a scalar and the last term in the
r.h.s. of the Schroedinger equation \SCHRXP\ drops out. The problem
appears to be equivalent to the quantum mechanical system of $N$
non-interacting fermions (due to the antisymmetry of $\phi(x)$) in a
potential $V(x)$. It was used in many applications, including the
solution of the non-critical string theory in 1+1 dimensions \KazMig .

The next smallest representation is adjoint. The adjoint wave function
satisfying the relation \TRANS\ should be a function of the type
$$\Psi(M;x)= \sum_{a=0}^{N-1} C_a(x) M^a$$ where the coefficients
$C_A$ possibly depend on the invariants (eigenvalues). If we denote
$\phi_{adj}(x_i;x)\equiv \phi_i(x)$ (depending of course on all $N$
$x_i$) we can write the Schroedinger equation for the adjoint wave
function in the form
\MARON :
\eqn\SADJ{ \sum_i\big(-\left({\p\over \p x_i}\right)^2+V(x_i)\big)\phi_k(x)
-{1\over N^2}\sum_{i(\ne k)} {\phi_i(x)-\phi_k(x)\over (x_i-x_k)^2}=
E\phi_k(x) }
One can see that the last term in the l.h.s. of this equation is $\sim
N^2$ smaller than the other terms and can be regarded as a small
perturbation on the background of the free fermion solution of the
singlet sector.

For one of physically most interesting applications, the 1+1
dimensional string theory, we need to solve the model in the inverted
oscillatorial potential $V(M)=-M^2$. The model is unstable and one
needs to specify the boundary conditions for big $M$'s.  Usually one
considers the boundary conditions when the absolute value of any of
the eigenvalues of $M$ cannot exceed some maximum value $\Lambda$ (a
cut-off wall). In the case of the large $N$ limit one takes $\Lambda
\sim N$ and it happens that the spectrum density of the model depends
in a very universal (logarithmic) way on $\Lambda$. In the singlet
state the spectrum is that of $N$ independent fermions (eigenvalues)
in the same potential and the eigenfunctions are the Slater
determinants of the parabolic cylinder functions (see the review in
\KAZDO . In the non-singlet sectors the eigenvalues start interacting
and obey a more complicated statistics corresponding to the symmetry
of the Young tableau of representation (see the review \POLY\ for the
details). Although the problem is clearly integrable the spectrum of
the non-singlet sectors of the inverted matrix quantum oscillator is
still unknown (for the large N estimates of the mass gap of adjoint
representation see \MARON ,\KLGR\ and \KAZBU).

It was conjectured in \KLGR\ and shown in \KAZBU\ that the adjoint
representation describe the vortex anti-vortex sector in the 1+1
dimensional string theory with one compact dimension. Higher
representation describe higher numbers of vortex anti-vortex pairs
(corresponding to the number of boxes in the Young tableau of the
representation).

\newsec{Character expansion and new solvable (multi) matrix models}

The group character expansion has shown its power in the lattice gauge
theory long time ago, starting from the work of A. Migdal \MIG .

The character expansion method proposed in the papers \KSWI - \KSWIII\
and inspired by the result of paper \DIFZ\ is the most general
approach for the reduction of the number of degrees of freedom from
$\sim N^2$ to $\sim N$ in a new big class of (multi) matrix
integrals. The matrix integral considered in these papers looks as
follows:
\eqn\DWG{  Z=\int \dN M \exp[-\Tr M^2 + \Tr V(AM)] }
where $V(y)=\sum_{k>2} t_k y^k$ is an arbitrary potential and A is an
arbitrary hermitian matrix (which can be taken diagonal without a loss
of generality). We again diagonalize the matrix $M$ as
\eqn\DECO{M=\O^+X\O}
The integral over the $U(N)$ variable $\O$ looks difficult to do
directly since the Itzykson-Zuber formula \CCIZ\ seems to be of little
use here. Instead of it let us expand $\exp[ \Tr V(AM)]$ as an
invariant function of the variable $AM$ in terms of the characters
$\chi_R(AM)$ of irreducible representations $R$ of the $GL(N)$ group:
\eqn\CHEX{  \exp[ \Tr V(AM)]=\sum_R f_R \chi_R(AM)  }
where the coefficients $f_R$ are the functions of $N$ highest weight
components of a representation 
$$R=\{0\le m_N\le m_{N-1}\le \cdots \le m_1<\infty\}.$$ The sum
$\sum_R$ is nothing but the sum over N ordered integers.  They can be
calculated due to the orthogonality of characters as the following
unitary integrals:
\eqn\COEF{ f_R=\int [d\O]_{U(N)}  \exp[ \Tr V(\O)] \chi_R(\O^+)  }
Note that this integral can be represented as an explicit integrals
only over the Cartan subgroup $\O=\{e^{i\o_1},\cdots,e^{i\o_N}\}$ and
thus contains only $N$ integration variables. We have $\Tr
V(\O)=\sum_k V(e^{i\omega_k})$ and $[d\O]_{U(N)} \ra \prod_k d\theta_k
\prod_{i>j}\sin^2{\theta_i-\theta_j\over 2}$. Now if we plug \CHEX\ into
\DWG\ we realize that the decomposition \DECO\ is actually useful and
we can integrate over $\O$ using the following orthogonality relation
between matrix elements of representation $R$:
\eqn\ORTH{ \int [d\O]_{U(N)}   \chi_R(A\O^+X\O)= {1\over {\it dim}_R  }
\chi_R(A)\chi_R(X)     }
where ${\it dim}_R$ is the dimension of a representation R.  We see
that we achieved our main goal: due to the formulas
\CHEX , \COEF\ and \ORTH\ we reduced  the original matrix 
integral \DWG\ to an integral over only $N$ eigenvalues
$x_1,\cdots,x_N$ of the matrix $M$ and the sum over $N$ highest weight
components $m_1,\cdots,m_N$.  In the large N limit, if we scale
appropriately the constants in the potential $V(M)$, the sums over
$m$'s can be replaced by integrals and we can again apply the saddle
point approximation in all $2N$ integration variables.  To get
explicitly the right large $N$ scaling of the couplings one usually
changes $e^{V} \ra e^{N V}$. Then the effective action at the saddle
point is always of the order $1/N^2$ and the new couplings of the
potential $V$ can be kept finite in this limit.

As was shown in \DIFZ\ (see also \KSWI ), the integral over
$x_1,\cdots,x_N$ can be calculated exactly and the remaining sum over
strictly ordered nonnegative integers $h_i=-m_i+N-i$ (shifted highest
weights) reads:
\eqn\ITZD{ Z=\sum_{h_1<h_2<\cdots<h_N} 
{\prod (h^e-1)!! h^o!! \over \prod (h^e-h^o)} \chi_R(A)\chi_R(t) }
where $\{h^e\}$ and $\{h^o\}$ are the collections of even and odd
 integers $h_k$ (their number is equal). Only the representations with
 equal amounts of even and odd $h$'es contribute to \ITZD . The
 products in the numerator go over all even and odd $h$'es and the
 product in the denumenator goes over all couples $h^e,h^o$.
 $\chi_R(t)$ is a character of the coupling constants $t_k$ written in
 the Schur form:
\eqn\Schur{\chi_R=det_{ij} P_{h_i-j}(t) }
and the Schur polynomials $P_k(t)$ are defined as usually: $\sum_n
P_n(t)z^n= e^{\sum_k t_kz^k}$. 

So in the large $N$ limit we have to do the saddle point calculation only 
with respect to $N$ summation variables $h_1,\cdots,h_N$.

The details of these formulas can be found in \DIFZ , \KSWI -\KSWIII .
One can also find in these papers the geometrical interpretation of
the integral \DWG\ in terms of the so called dually weighted planar
graphs. It gives the generating function of planar graphs where both
vertices and faces are weighted by the generating parameters depending
on their orders.  In \KSWI -\KSWIII\ one can find the solutions of
some combinatorial problems related to the enumeration of planar
graphs which were possible only due to the power of the character
expansion method. The particular solutions of the saddle point
equations could be very tricky but it is already a ``classical''
problem of solution of various integral equations rather than a
``quantum'' problem of functional integration over infinite
matrices. In that sense this model is solvable. 

It is obvious that there exist many ways to generalize the model \DWG\
to other matrix integrals. An immediate generalization is to
substitute the $\Tr M^2$ term in \DWG\ by an arbitrary function
$W(M)$.  In that case we cannot calculate explicitly the coefficients
$f_R$ (except  when $W$ is a monomial: $W(M)=M^k$) but we still get an 
explicit integral over $3N$ variables $x_i$, $\o_i$ and $m_i$. So the model 
is again solvable.

Another solvable matrix model of this kind involving general complex
matrices was proposed and investigated in \KS , \KSTW . Its free
energy gives a generating functional counting branched coverings of
two dimensional surfaces.

The most general solvable two matrix model reads as
\eqn\GTWM{ Z=\int \ \dN A \ \dN B\exp N\Tr
 \big(U(AB)+V(A)+ W(B)\big)}
where $U$,$V$ and $W$ are arbitrary functions. The way to reduce it to
$\sim N$ degrees of freedom is again to expand in characters
\eqn\CHEXU{  \exp[ \Tr U(AB)]=\sum_R u_R \chi_R(AB)  }
diagonalize the matrices $A$ and $B$ and integrate over the  
$U(N)$ variable between them by means of \ORTH . In a particular case 

 \eqn\TWMF{ Z=\int \dN A \ \dN B\exp N\Tr
 \big( {1\over 2}(A^2+ B^2)-{\alpha\over 4}( A^4+
B^4)-{\beta\over 2} (AB)^2\big)}
the model describes a special trajectory of the 8-vertex model on
random graphs. It was completely solved in \KZJ . Again it was
possible, using character orthogonality relations, to integrate over
the relative angle between $A$ and $B$; this leads to  separation
into one-matrix integrals:
\eqn\charexp{
{\rm Z}(\alpha,\beta)\sim \sum_{\{ h\}} (N\beta/2)^{\# h/2} c_{\{h\}} 
[P_{\{ h\}}(\alpha)]^2}
where $c_{\{ h\}}$ is a coefficient:
$$c_{\{ h\}}={1\over\prod_i \lfloor h_i/2\rfloor ! \prod_{i,j}
(h_i^{\rm even}-h_j^{\rm odd})}$$ and $S_{\{ h\}}(\alpha)$ is a
one-matrix integral
\eqn\defR{
S_{\{ h\}}(\alpha)=\int \dN M\, \chi_{\{ h\}}(M) \exp N\left[-{1\over
2}\tr M^2 +{\alpha\over 4}\tr M^4\right] }
which appears squared in \charexp\ because the contributions from the
two matrices $A$ and $B$ are identical.

Now we can reduce the calculation of the one-matrix integral
$S_{\{h\}}$ to  eigenvalue integrations:
\eqn\eigenR{
S_{\{ h\}}(\alpha)=\int \prod_k d\lambda_k\, \Delta(\lambda)
\det\left(\lambda_k^{h_j}\right) \exp N\left[
-{1\over 2}\sum_k \lambda_k^2+{\alpha\over 4}\sum_k \lambda_k^4\right]}
where $\Delta(\lambda)=\det\left(\lambda_k^{N-j}\right)=
\prod_{j<k}(\lambda_j-\lambda_k)$.

Now we are left only with  $N$ degrees of freedom and the action of the
order $N^2$, so the integration  is reduced to the saddle point calculation
with respect to the eigenvalues $\lambda_k$ (see \KZJ\ for the details).

We can immediately propose some solvable generalization of the 
general two matrix model \GTWM\ to a multi-matrix chain:
 \eqn\MMM{ Z=\int\prod_{q=1}^Q \dN M_q\exp \Tr\big(\sum_{q=1}^Q V_q(M_q) 
+\sum_{p=1}^{Q-1} W(M_{p-1}M_p) \big)}
where $V$ and $W$ are arbitrary functions. 

Another interesting model solvable by the character expansion method
can be written in the following general form:
 \eqn\MMM{ Z=\int\prod_{q=1}^Q \dN M_q\exp \big(\Tr V_q(M_q) 
+V(\prod_{p=1}^{Q}M_p) \big) }
It can be solved by character expansion with respect to the last
factor by means of the formulas \CHEX ,\COEF\ and the multiple
application of the formula \ORTH\ by induction.

The results in terms of the sum over $N$ highest weight components of
the representations $R$ reads:
\eqn\PPPM{
 Z= \sum_R {f_R\over [{\it dim}_R]^{Q-1} }  
\prod_{q=1}^Q[S^{(q)}_{\{ h\}}]        }
where
\eqn\SSSM{
S^{(q)}_{\{ h\}}=
\int \dN M\, \chi_{\{ h\}}(M) \exp\Tr V_q(M)     }
The last integral can be immediately reduced to the integrations of a
type \eigenR\ over eigenvalues of the matrix $M$.

\newsec{ Comments and unsolved problems}

A few comments are in order:

1. The non-singlet sectors in the matrix quantum mechanics \MQM\ can be
   effectively studied for the oscillatorial potential $V(M)= M^2$. In
   this case the Hamilton-Ian is a collection of $N^2$ independent
   oscillators represented by the matrix elements of $M$.  The
   spectrum of Hamilton-Ian of this model in a given irreducible
   representation of $U(N)$, is encoded into the partition functions
   $Z_R(q)$ for finite inverse temperature $\beta$ (where $q=e^\beta$)
   in a given representation $R$. The effective way to study $Z_R(q)$
   can be found in papers \KAZBU\ or \HKK .

2. The character expansion is nothing but the Fourier expansion on a
   group manifold. As trivial as it looks for us now the Fourier
   transform was always a powerful method of solving problems using
   their symmetries. Many of the matrix models presented in the
   previous section and solved by this method seemed hopeless just a
   few years ago.

3. One of the interesting and not well studied questions is how to
   classify all the matrix integrals which can be reduced from $\sim
   N^2$ to $\sim N$ integrals or sums by use of the character
   expansion.

4. In many physically interesting cases we don't need a general form
   of potentials $V(M)$ or $W(M)$ mentioned through this paper. For
   example, as it was mentioned, to study the universal behavior of
   the large $N$ matrix quantum mechanics near the instability point
   we need to know only the solution in the vicinity of a quadratic
   top of the potential $V(M)\simeq -\Tr(M-M_0)^2$. The rest of the
   potential $V(M)$ has small influence on the behavior of the
   eigenvalues and serves only as a $U(N)$ invariant cutoff wall. It
   simplifies greatly the problem. For instance, all the applications
   in string theory, two-dimensional quantum gravity and most of
   statistical-mechanical applications need only the analyses of the
   vicinity of such critical points. The lesson to draw from it is
   that for some physically most interesting regimes the seemingly
   hopeless matrix integrals become not so hopeless and look ``almost
   Gaussian''. May be a general method of the investigation of these
   instability points can be worked out.

5. Another question is related to the integrability properties of sums
   and integrals after such reduction. The partition functions of some
   of them (such as the old one matrix and two matrix models) are
   known to be $\tau$-functions of some integrable hierarchies of
   classical differential or difference equations, like Toda hierarchy
   \MMM , \HKK\ or KP hierarchy \KKN\ (see \KOSR for a good
   introduction).  But many others, like the model \DWG , cannot be
   represented by free fermions. On the other hand, the
   Itzykson-DiFrancesco formula \ITZD\ suggests that it might exist
   some interacting fermion representation of the partition function
   of the model of dually weighted graphs \DWG .

The method of character expansion as well as all other methods of
calculation of the large $N$ matrix integrals presented here represent
just another refining and generalization of the usual method of
reduction to the matrix eigenvalues invented long ago by Dyson. Its
range of applicability is quite limited although it includes quite a
few important matrix integrals known from physics and mathematics.
Many more interesting matrix integrals look not hopeless for the
investigation in the large $N$ limit.  The search for new tricks of
integration over matrices is a fascinating and potentially extremely
rewarding research direction.

\newsec{Acknowledgments}
%
\noindent  
I am grateful to the organizers of the MSRI Workshop ``Matrix models
and Pinlev\'e equations'' P.~Bleher and A.~Itz for the kind
hospitality and fruitful discussion during the Workshop.

I also would like to thank I.~K.~Kostov for useful comments and
M.~Fukuma for the illuminating discussions concerning the matrix
quantum mechanics in adjoint representation.


\listrefs

\bye